\documentclass[11pt]{cernrep}
\usepackage{graphicx}
\usepackage{amsmath}
\usepackage{amsfonts,amssymb}
\usepackage{here}
\begin{document}
 \title{Coupled-cluster renormalization group}
\author{Amir H. Rezaeian and Niels R. Walet}
\institute{Department of Physics, UMIST, PO Box 88, Manchester, M60 1QD, UK}
\maketitle
\begin{abstract}
The coupled cluster method (CCM) is one of the most successful and
universally applicable techniques in quantum many-body theory.  The
intrinsic nonlinear and non-perturbative nature of the method is
considered to be one of its advantages.  We present here a combination
of CCM with the Wilsonian renormalization group which leads to a powerful
framework for construction of effective Hamiltonian field theories.  As
a toy example we obtain the two-loop renormalized $\phi^{4}$ theory.
\end{abstract}

\section{INTRODUCTION}
The CCM originated in nuclear physics around forty years ago on the
work of Coester and K\"{u}mmel \cite{1}. There exists several versions
of CCM formulation \cite{2}. They are denoted generically as
independent-cluster (IC) parametrizations, in the sense that they
incorporate the many-body correlations via sets of amplitudes that
describe various correlated clusters within the interacting system as
mutually independent entities. The IC methods differ in the way they
incorporate the locality and separability properties.  Each of the IC
methods has been shown to provide an exact mapping of the original
quantum mechanical problem to a corresponding classical mechanics in
terms of a set of multiconfigurational canonical field amplitudes. In
this way one can introduce a particular differentiable IC manifold
which is endowed with a symplectic structure. The basic IC amplitudes
provide the local coordinates on these manifold.

Despite extensive studies of the CCM in different areas of physics,
limited progress has been made in application of the CCM to
modern particle physics \cite{niels}. In the recent paper \cite{3, 4}
we showed that the implementation of the Wilsonian exact
renormalization group approach in the CCM framework leads to an
elegant method for obtaining Hamiltonian renormalization flows.  Our
method resembles the similarity renormalization group \cite{g} (see
the talk by Glazek), since we employ a double non-unitary transformation
(using the extended CCM) to decouple the irrelevant degrees of
freedom, which leads to partial diagonalization of the Hamiltonian. In
this note we present a short review of our approach.

\section{FORMALISM}
 Notice that our formulation does not depend on the form
of the dynamics (i.e. equal time or light cone).
 We assume that generally the renormalized Hamiltonian
$H^{\text{eff}}(\lambda)$ up to scale $\lambda$ can be expressed
as
\begin{equation}
H^{\text{eff}}(\lambda)=H(\lambda)+H_{C}(\lambda), \label{oh}
\end{equation}
where $H_{C}(\lambda)$ is a {}``counterterm''. Now we introduce a large cut-off $\Lambda$. We define
two subspaces, the model-space $P:\{|L\rangle
\bigotimes|0,b\rangle_{h}, L\leq \mu\}$ and the complement-space $Q:
\{|L\rangle \bigotimes \left(|H\rangle-|0,b\rangle_{h}\right),\mu<H\leq\Lambda \}$. The ket
$|0,b\rangle_{h}$ is the bare high-energy vacuum (the ground state of
the free high-momentum Hamiltonian). The $P$-space contains
interacting low-energy states and the $Q$-space contains the orthogonal
complement (the symbols $|L\rangle$ and $|H\rangle$ denote generic
low- and high-energy states, respectively). Our renormalization
approach is based on decoupling of the complement space $Q$ from the
model space $P$ by using a non-unitary transformation aiming at
constructing of the counterterms. The transformation of $H(\Lambda)$ is
defined by
\begin{equation}
\overline {H}=e^{\hat{S}'(\mu,\Lambda)}e^{-\hat{S}(\mu,\Lambda)}H(\Lambda)e^{\hat{S}(\mu,\Lambda)}e^{-\hat{S}'(\mu,\Lambda)}\equiv H(\mu)+\delta H(\mu,\Lambda),
\end{equation}
and we can expand $\hat{S}(\hat{S}')$ in terms of the independent
coupled cluster excitation $I$, assuming a naive power counting, 
\begin{eqnarray}
&&\hat{S}=\sum_{m=0}\hat{S}_{m}\left(\frac{\mu}{\Lambda}\right)^{m}, \hspace{2cm} \hat{S}_{m}=\sideset{}{'}\sum_{I}\hat{s}_{I}C^{\dag}_{I},\nonumber\\
&&\hat{S}'=\sum_{m=0}\hat{S}'_{m}\left(\frac{\mu}{\Lambda}\right)^{m}, \hspace{2cm} \hat{S}'_{m}=\sideset{}{'}\sum_{I}{}\hat{s}'_{I}C_{I}.\label{s}\
\end{eqnarray}

Here the primed sum means that $I\neq 0$, and momentum conservation is
included in $\hat{s}_{I}$ and $\hat{s}'_{I}$. The $C_{I}$ and
$C^{\dag}_{I}$ are annihilation and creation operators in the
high-energy Fock space for the chosen quantization scheme.  The indices
${I}$ define a subsystem, or cluster, within the full system of a
given configuration. Notice that $\hat{s}_{I}$ and $\hat{s}'_{I}$ are
parameters in the high-energy Fock space, while they are
operators in the low-energy Fock space.  Therefore one can guarantee
the proper size-extensivity and conformity with the linked-cluster theorem
(at any level of approximation) in the Wilsonian high-energy
shell. However one can still apply the standard CCM parametrization after
obtaining the effective low-energy Hamiltonian (thus the
size-extensivity can be extended to the whole Fock space).

It is well-known in CCM formulation that the parametrization
Eq.~(\ref{s}) is compatible with the Hellmann-Feynman theorem and the
phase space $
\{\hat{s}_{I},\hat{s}'_{I}\}$ for a given $m$ is a symplectic
differentiable manifold. Having said that, the representation in this
way can no longer remain manifestly hermitian. This in fact give rise
to some superfluous degrees of freedom since the actual phase space is
enlarged into a complex manifold. There is long tradition for such
approaches, of course with different motivation (e.g., in the BRST
formalism, the phase space is expanded by anti-commuting canonical
coordinates). As Arponen emphasized \cite{5} the most important reason
behind the extra degrees of freedom in the CCM is the fact that one
can introduce a universal average value functional allowing the
simultaneous computation of the expectations of other operators than
the Hamiltonian.  In Ref.~\cite{4} we showed that this non-unitarity
leads to an economic computation and does not induce any relevant
non-hermiticity in the renormalization group sense. On the other hand
this can, in principle, lead to Poincar\'{e} invariance at any given
level of truncation regardless of the regularization scheme. One may
now impose the decoupling conditions derivable from the dynamics of
the quantum system, leading to diagonalization of the transformed
Hamiltonian matrix (double similarity transformation splits
diagonalization to upper and lower triangle part) ,
\begin{eqnarray}
&&Q\overline{H}P=0 \to \langle 0|C_{I}e^{-\hat{S}}He^{\hat{S}}|0\rangle=0, \label{eq17}\\
&&P\overline{H}Q=0 \to \langle 0|e^{\hat{S}'}e^{-\hat{S}}He^{\hat{S}}e^{-\hat{S}'}C^{\dag}_{I}|0\rangle=0,\label{eq19}\
 \end{eqnarray}
where $I\neq 0$. Thus the effective low-energy Hamiltonian is
\begin{equation}
\hat{H}^{\text{eff}}=P\bar{H}P \equiv ~_{h}\langle
b,0|e^{\hat{S}'(\mu,\Lambda)}e^{-\hat{S}(\mu,\Lambda)}H(\Lambda)e^{\hat{S}(\mu,\Lambda)}e^{-\hat{S}'(\mu,\Lambda)}|0,b\rangle_{h}.
\label{eq15}
\end{equation}
The decoupling conditions Eqs.~(\ref{eq17},\ref{eq19}) make the $P$
sector of the truncated Fock space independent of the rest. This means
that the contribution of the excluded sector of Hilbert space (the
high-energy space) is taken into account by imposing the decoupling
conditions. These are sufficient requirements to secure partial
diagonalization of the Hamiltonian in the fast-particle space. This
technique seems to be more universal than Wilsonian renormalization
based on Lagrangian framework since one can eliminate other
irrelevant degrees of freedom in a unified way. 

Here, by construction, the energy-dependent Bloch-Feshbach formalism
is made free of the small-energy denominators which plague
perturbation theory. One can then determine the counterterm by
requiring coupling coherence \cite{6}, namely that the transformed
Hamiltonian Eq.~(\ref{eq15}) has the same form given in
Eq.~(\ref{oh}), with $\lambda$ replaced by $\mu$. By means of the
decoupling conditions one can determine the unknown individual
amplitudes $\hat{s}_{I}$ and $\hat{s}'_{I}$ for a given $m$ in a
consistent truncation scheme which is the so-called SUB($n$,$m$). The
choice of $n$ depends on the bare Hamiltonian interaction and should
be fixed from the outset.

The formulation is intrinsically non-perturbative, however we can
simply apply a perturbation expansion.  We split the Hamiltonian in five
parts:
$H=H_{1}+H_{2}^{\text{free}}(\text{high})+V_{C}(C_{I}^{\dag})+V_{A}(C_{I})+V_{B}$
where $H_{1}$ contains only the low frequency modes with $k\leq\mu$,
$H_{2}$ is the free Hamiltonian for all modes with $\mu<k<\Lambda$,
$V_{C}$ contains low frequency operators and products of the high
frequency creation operators $C_{I}^{\dag}$ and $V_{A}$ is the
hermitian conjugate of $V_{C}$. The remaining terms are contained in
$V_{B}$. If we assume that $V_{A,C,B}$ are of first order in coupling
constant, then one can expand Eqs.~(\ref{eq17},\ref{eq19}) in order of
$m$, leading to the elimination of the fast-particle up first
order in the coupling. This produces an effective Hamiltonian up to
third order in the coupling:
\begin{eqnarray}
 &  & m=0:\langle0|C_{I}(V_{C}+[H_{2},\hat{S}_{0}])|0\rangle=0, \nonumber\\
 &  &~~~~~~~~~~:\langle0|(V_{A}-[H_{2},\hat{S}'_{0}])C_{I}^{\dag}|0\rangle=0,  \nonumber \\
 &  & m=1:\langle0|C_{I}([H_{1},\hat{S}_{0}]+[H_{2},\hat{S}_{1}]+[V_{A},\hat{S}_{1}]+[V_{C},\hat{S}_{1}])|0\rangle=0,\nonumber \\
 &  &~~~~~~~~~~:\langle0|([H_{1},\hat{S}'_{0}]+[H_{2},\hat{S}'_{1}]+[V_{C},\hat{S}'_{1}]+[V_{A},\hat{S}'_{1}]-[V_{A},\hat{S}_{1}])C_{I}^{\dag}|0\rangle=0,\nonumber \\
 &  & \hspace{2cm}\vdots\nonumber \\
 &  & m=n:\langle0|C_{I}([H_{1},\hat{S}_{n-1}]+[H_{2},\hat{S}_{n}]+[V_{A},\hat{S}_{n}]+[V_{C},\hat{S}_{n}])|0\rangle=0,\nonumber \\
 &  &~~~~~~~~~~~:\langle0|([H_{1},\hat{S}'_{n-1}]+[H_{2},\hat{S}'_{n}]+[V_{C},\hat{S}'_{n}]+[V_{A},\hat{S}'_{n}]-[V_{A},\hat{S}_{n}])C_{I}^{\dag}|0\rangle=0. \label{eq21}\
\end{eqnarray}

\section{EXAMPLE}
We now apply this formalism to the computation of the effective
Hamiltonian for $\phi^{4}$ theory up to two-loop order in
equal-time quantization. The details can be found in Ref.~\cite{4}. The bare $\phi^{4}$ theory Hamiltonian is
\begin{equation}\label{a1}
H=\int d^{3}x\left(
\frac{1}{2}\pi^{2}(x)+\frac{1}{2}\phi(x)\big(-\nabla^{2}+m^{2}\big)\phi(x)+g\phi^{4}(x)\right).
\end{equation}
We now split field operators into high- and low-momentum modes;
$\phi(x)=\phi_{L}(x)+\phi_{H}(x)$, where $\phi_{L}(x)$ denotes modes
of low-frequency with momentum $k\leq \mu$ and $\phi_{H}(x)$ denotes
modes of high-frequency with momentum constrained to a shell
$\mu<k\leq\Lambda$. The $\phi_{H}(x)$ is represented in the Fock space
as
$\phi_{H}(x)=\sum_{\mu<k\leq\Lambda}\frac{1}{\sqrt{2\omega_{k}}}(a_{k}e^{ikx}+a^{\dag}_{k}e^{-ikx})$. One
can plug this expansion into Eq.~(\ref{a1}) and produce the Fock
representation of the Hamiltonian (in the high-energy space), the
high-energy configurations in the Fock space are specified by
\(\{C_{I}\to \prod_{i=1} a_{k_{i}}\}\) and \(\{C^{\dag}_{I}\to
 \prod_{i=1}a_{k_{i}}^{\dag})\}\). 
Up to two-loop expansion, our renormalization scheme requires us to keep
$S(S')$ at least to order $n=4$. The $\hat{S}(\hat{S})$ operators
consistent with a $SUB(4,m)$ truncation scheme are,
\begin{eqnarray}
\hat{S}_{m}&=&\int\sum\left(\hat{S}^{1}_{m}~a^{\dag}_{k}+\hat{S}^{2}_{m}~a^{\dag}_{k}a^{\dag}_{p}+ \hat{S}^{3}_{m}~a^{\dag}_{k}a^{\dag}_{p}a^{\dag}_{q}+\hat{S}^{4}_{m}~a^{\dag}_{k}a^{\dag}_{p}a^{\dag}_{q}a^{\dag}_{r}\right),\nonumber\\
\hat{S}'_{m}&=& \int\sum \left(\hat{S}'^{1}_{m}~a_{k}+\hat{S}'^{2}_{m}~a_{k}a_{p}+\hat{S}'^{3}_{m}~a_{k}a_{p}a_{q}+\hat{S'}^{4}_{m}~a_{k}a_{p}a_{q}
a_{r}\right).\label{a6}\
\end{eqnarray}
The unknown coefficients $\hat{S}_{m}(\hat{S}'_{m})$ in Eq.~(\ref{a6}) can be found by Eqs.~(\ref{eq17},\ref{eq19}) or Eq.~(\ref{eq21}).
We restrict ourselves to the elimination
of the high-energy degrees of freedom up to first order in the
coupling constant $g$ and in a consistent truncation scheme $\text{SUB}(4,2)$. For $m=0$, the potential divergent terms arise from 
\begin{eqnarray}
\delta H &=& \langle 0|H+[H,\hat{S}^{2(3)}_{0}]+[[H,\hat{S}_{0}],\hat{S}'_{0}]|0\rangle=\frac{3g}{4\pi^{2}}(\Lambda^{2}-\mu^{2})\int d^{3}x \phi^{2}(x)\nonumber\\
&-&\frac{9g^{2}}{2\pi^{2}}\ln\left(\frac{\Lambda}{\mu}\right)\int d^{3}x \phi^{4}(x)-\frac{3g^{2}}{2\pi^{4}}(2\ln 2-1)\Lambda^{2}\int d^{3}x \phi^{2}(x)\nonumber\\
&+&\frac{3g^{2}}{16\pi^{4}}\ln\left(\frac{\Lambda}{\mu}\right)\int d^{3}x(\nabla\phi(x))^{2}
+\frac{27g^{3}}{2\pi^{4}}\Big[\big[\ln\left(\frac{\Lambda}{\mu}\right)\big]^{2}+\ln\left(\frac{\Lambda}{\mu}\right)\Big]\int d^{3}x. \
\end{eqnarray}
One can immediately deduce the renormalization factors $Z_{m}$, $Z_{g}$ and $Z_{\phi}$ from the above equation,
\begin{eqnarray}
Z_{m}&=&1-\frac{3g}{2\pi^{2}}(\Lambda^{2}-\mu^{2}), \hspace{2cm} Z_{\phi}=1-\frac{3g^{2}}{8\pi^{4}}\ln\left(\frac{\Lambda}{\mu}\right).\nonumber\\
Z_{g}&=&1+\frac{9g^{2}}{2\pi^{2}}\ln\left(\frac{\Lambda}{\mu}\right)+\frac{g^{3}}{4\pi^{4}}\left(81\left(\ln\left(\frac{\Lambda}{\mu}\right)\right)^{2}-51\ln\left(\frac{\Lambda}{\mu}\right)\right). \label{z}\
\end{eqnarray}
In the same way one can obtain the unknown coefficients for $m=1$. The only divergent contribution up to order $g^{2}$ arises from
\begin{equation}
\delta H=-\langle 0|[H_{1},\hat{S}_{1}],\hat{S}'_{0}]|0\rangle=-\frac{3g^{2}}{16\pi^{4}}\ln\left(\frac{\Lambda}{\mu}\right)\int d^{3}x  \pi^{2}(x),\label{pi}
\end{equation}  
which contributes to the two-loop wave-function renormalization $Z_{\pi}$, namely it leads to $Z_{\pi}=Z_{\phi}^{-1}$. One now may reproduce the well-known two-loop $\beta$- and $\gamma$-functions by making use of Eq.~(\ref{z}).
\section{CONCLUSION}
In this note we have outlined a strategy to construct effective Hamiltonian
field theories in framework of well-known coupled cluster many body
theory. We showed that this method has several advantages: 1: all
other irrelevant degrees of freedom like many-body states can be
systematically eliminated in the same way, 
2: renormalization flows can be obtained perturbatively or non-perturbatively.
3: the formulation
is free of any small-energy denominator plaguing old-fashion perturbation theory, 
4: with the implementation of the decoupling
properties, having the phase space extended to a complex manifold, one can in
principle preserve Poincar\'{e} invariance regardless of a regularization scheme,
5: the formulation is not restricted to any quantization scheme (e.g. equal time or light cone).

\section*{ACKNOWLEDGEMENTS}
One of the authors (AHR) acknowledges support from British Government
ORS award. The work of NRW is supported by the EPSRC.

\end{document}